\documentclass{elsart}
\usepackage{epsfig,bbm}
\journal{Nuclear Physics B}
\begin{document}

\begin{frontmatter}
\title{Liquid-gas and other unusual thermal phase 
transitions in some large-$N$ magnets}
\author{O. Tchernyshyov}, 
\ead{otcherny@princeton.edu}
\author{S. L. Sondhi} 
\ead{sondhi@feynman.princeton.edu}
\address{
Department of Physics, Princeton University, Princeton, NJ 08544, USA} 

\begin{abstract}
Much insight into the low temperature properties of quantum magnets
has been gained by generalizing them to symmetry groups of order $N$,
and then studying the large $N$ limit. In this paper we consider an
unusual aspect of their finite temperature behavior---their exhibiting
a phase transition between a perfectly paramagetic state and a
paramagnetic state with a finite correlation length at $N = \infty$.
We analyze this phenomenon in some detail in the large ``spin''
(classical) limit of the SU$(N)$ ferromagnet which is also a lattice
discretization of the CP$^{N-1}$ model. We show that at $N = \infty$
the order of the transition is governed by lattice connectivity.  At
finite values of $N$, the transition goes away in one or less
dimension but survives on many lattices in two dimensions and higher,
for sufficiently large N. The latter conclusion contradicts a recent
conjecture of Sokal and Starinets \cite{SS}, yet is consistent with
the known finite temperature behavior of the SU$(2)$ case.  We also
report closely related first order paramagnet-ferromagnet transitions
at large $N$ and shed light on a violation of Elitzur's theorem at 
infinite $N$ via the large $q$ limit of the  $q$ state Potts model, 
reformulated as an Ising gauge theory.
\end{abstract}

\begin{keyword}
$1/N$ expansion \sep
quantum magnetism \sep
nonlinear $\sigma$ model \sep
CP$^{\rm N-1}$ model \sep
phase transitions
\end{keyword}

\end{frontmatter}


\section{Introduction}
\label{section-intro}

The properties of quantum antiferromagnets in low dimensions have been
intensely studied over the past decade and a half. Much insight has
been gained by large $N$ treatments based on generalizing the symmetry
group from SU(2) to either SU($N$) or Sp($N$), especially in two
dimensions where exact solutions are not available. These
reformulations involve the representation of the spins by bilinears in
fermionic or bosonic ``spinon'' operators at the cost of introducing
local constraints on their number and an associated gauge invariance
\cite{AM,AA,Sachdev}.  At $N=\infty$ the constraints are trivially
solved and a purely quadratic problem results. Most of this effort has
gone into elucidating the zero temperature phase diagram and has also
involved going beyond the $N=\infty$ limit by thinking about the
structure of the gauge theory that results if the spinons are
integrated out.

An oddity from the early work using bosons is the report \cite{AA}, on
a square lattice, of a finite-temperature {\it phase} at $N=\infty$
with no intersite correlations---a perfect paramagnet.  At low
temperatures the energy cost compels spins to align (in ferromagnets)
or antialign (in antiferromagnets).  In the language of Schwinger
bosons, this is seen as a correlation or anticorrelation of boson
flavors on adjacent sites.  At higher temperatures, free energy
$F=E-TS$ is dominated by entropy.  When the number of boson flavors
$N\to\infty$ (with the coupling constant appropriately rescaled $J 
\mapsto J/N$), the entropy of the disordered state completely
overpowers the energy cost, so that neighboring sites become {\em
perfectly} uncorrelated above a certain temperature.  (At large but 
finite $N$ the high-temperature phase has nonvanishing correlations.)
Such a phase must then be separated from the lower temperature
paramagnetic state with finite intersite correlations by a phase
transition. This phase transition clearly has no analog in the
physical SU$(2)$ problem and is therefore an embarrassment for the
large-$N$ approach.\footnote{A similar embarassment arises in the
infinite-$N$ solution of the Kondo problem which exhibits truly
non-analytic behavior at a finite temperature. There the exact Bethe
Ansatz solutions were used to show that this happens only at infinite
$N$; otherwise it is a crossover that sharpens continuously as $N$ is
increased \cite{Coleman,CA}.} We have found that an essentially
identical transition occurs in the SU$(N)$ generalization of the
Heisenberg ferromagnet at $N=\infty$.  Understanding its fate at
finite $N$ is equally a matter of interest.

In this paper we investigate this transition in some detail.  To make
life simpler we have restricted ourselves to the SU$(N)$ ferromagnet,
although much of what we say should apply {\it mutatis mutandis} to
the Sp$(N)$ antiferromagnet.  We make one further simplification, that
of taking the large ``spin'' or boson density limit at any fixed $N$,
which renders the problem classical without destroying the transition
of interest.\footnote{Much of what we have to say should go through at
large but not infinite ``spin''---at finite temperatures this
distinction is quantitative.}  For the classical SU$(N)$
ferromagnet,we first examine the $N=\infty$ solution carefully and
show that the transition is first order on the square lattice, a fact
which will be crucial in the following.  We analyze a number of other
lattices and find that this is the outcome on all lattices that have a
shortest closed loop of length three or four. On other lattices, such
as the honeycomb or the linear chain, which lack such loops, the
transition is continuous.  The order of the transition at $N=\infty$
is thus influenced by lattice connectivity and not
dimensionality. However, life does become more interesting in $d > 2$
where ferromagnetic states that break the SU$(N)$ symmetry become
stable to the Mermin-Wagner fluctuations contained in the $N=\infty$
theory.  Now it is possible for such states to ``piggy-back'' on the
finitely correlated paramagnets and first order transitions between
the perfect paramagnet and ferromagnetic states result in cases where
the ``underlying transition'' is predicted to be first order
``enough''.  The corresponding transition in the Sp$(N)$
antiferromagnet has been found recently by DeSilva and co-workers
\cite{DMZ}.

We turn next to the survival of this transition at finite values of
$N$. We offer strong evidence for the following conclusions.\\
(a) In $d=1$ it goes away, as it must on general grounds. \\
(b) In $d=2$ the first order transitions
survive for sufficiently large values of $N$ but terminate at an
Ising critical endpoint; the critical value $N_c$ is likely too large 
to be seen in feasible simulations. \\
(c) In $d=3$ and above the first order transitions again survive. In
cases where they are preempted by first order transitions to the
ferromagnetic state at $N=\infty$ the latter transition again survives
at large $N$ and presumably turns continuous before the SU$(2)$ limit 
is reached. \\
(d) In cases where the transition is continuous {\it at} $N=\infty$, we
conclude that the transition goes away at finite $N$.

The classical SU$(N)$ model we study has a pre-history for it is a
lattice version of the CP$^{N-1}$ model. Previous workers, most
notably Sokal and Starinets \cite{SS}, have concluded that the
transition in the lattice model is an artefact of $N=\infty$ in {\it
all} dimensions.  Their arguments are based on an exact solution of
the $d=1$ problem, a violation of gauge invariance at $N=\infty$ and
the apparent lack of such a transition at small $N$ in simulations. In
a companion paper to ours, Fendley and one of us (OT) have
independently solved the $d=1$ case and so we are in agreement with
Sokal and Starinets on that. We will argue below that the breakdown of
gauge invariance is misleading and in the course of this argument we
will appeal to a similar breakdown in an Ising gauge reformulation of
the $q$ state Potts model that exhibits a phase transition with a
family resemblance to the ones at issue and where one can appeal to
well-established results. We will argue that the transition at issue
has the character of a liquid-gas transition in that the two phases
can be smoothly continued into one another.  The two paramagnetic phases 
have different energy densities; the difference varies with $N$ and 
vanishes at some critical value $N_c$.  Finally, we will compute
the dimensionless surface tension at coexistence for the infinite-$N$
problem and show that it is small and thereby conclude that the
transition terminates at rather large values of $N$, consistent with
the failure to observe it in simulations.

In the balance of this section we introduce the quantum and classical
Hamiltonians of interest. In Section \ref{section-infiniteN} we carry
out a saddle point analysis at $N=\infty$. In Section
\ref{section-finiteN} we consider the finite (but large) $N$ problem. 
We summarize our conclusions in Section \ref{section-conclusion}.

\subsection{Hamiltonians}

The quantum problems we have in mind are the bosonic SU$(N)$ 
generalization of the Heisenberg antiferromagnet:
\begin{equation}
H = -\frac{J}{2N}\sum_{\langle ij \rangle} 
(b^\dagger_{i\alpha}b_{j\alpha})(b^\dagger_{j\beta}b_{i\beta})
\label{eq-H-quantum1}
\end{equation}
and the bosonic Sp$(N)$ generalization of the Heisenberg 
antiferromagnet:
\begin{equation}
H = -\frac{J}{2N}\sum_{\langle ij \rangle}
({\mathcal J}^{\alpha \beta} b^\dagger_{i\alpha} b^\dagger_{j\beta})
({\mathcal J}_{\gamma \delta} b_{i\gamma}b_{i\delta}).
\label{eq-H-quantum2}
\end{equation}
Here $\mathcal J$ is an antisymmetric block-diagonal matrix $2N \times 2N$:
\begin{eqnarray}
\mathcal J^{\alpha\beta} = 
\left(
\begin{array}
{rrrrrrrrrr}
0 & & 1\\
-1 & & 0\\
& & & & 0 & & 1\\
& & & & -1 & & 0\\
& & & & & & & & & \ddots 
\end{array}
\right).
\label{eq-eps-def}
\end{eqnarray}
The Greek indices run from $1$ to $N$ ($2N$ for the antiferromagnet)
and $\sum_{\alpha} b^\dagger_{i\alpha}b_{i\alpha} = n_b$ fixes the
``spin''.  For SU(2) $\equiv$ Sp(1), $n_b=2 S$ does fix the spin.  Note
that these generalizations, obtained by replacing the spin operators
by their bosonic ``square roots'' come with local constraints and
hence a local gauge invariance which we make more explicit below.

Considerable insight has been gained from considering the 
mean field theory that results in the $N \rightarrow \infty$
limit taken while keeping $\kappa = n_b/N$ fixed. However this 
limit also exhibits a finite temperature phase transition between
two paramagnetic phases, first discussed for the SU$(N)$
case by Arovas and Auerbach \cite{AA}.

To study this transition in more detail, we will study an easier
limit, that of $\kappa \rightarrow \infty$ at finite temperatures. 
With appropriately rescaled variables and units such that 
\[
b_{i\alpha} = z_{i\alpha} n_b/N, 
\hskip 1cm
J (n_b/N)^2 = 2, 
\]
this yields a classical partition function governed, in the SU$(N)$
case, by the classical energy
\begin{equation}
E = -\frac{1}{N}\sum_{\langle ij \rangle} |z^*_{i\alpha}z_{j\alpha}|^2,
\label{eq-E-classical}
\end{equation}
defined in terms of a complex $N$-vector ${z}$ obeying $z^\dagger
\cdot z=N$.  This Hamiltonian is invariant under global $SU(N)$
rotations, and also under the aforementioned local gauge
transformations $z_i \to e^{{\rm i}\alpha_i}z_i$ at any site $i$.
The vector $z$ takes values on a complex sphere $U(N)/U(N-1)$, which
as a manifold is identical to the real $2N-1$-sphere $O(2N)/O(2N-1)$.
However, the $U(1)$ gauge symmetry can be used to effectively reduce
the number of degrees of freedom in the problem by 1 by an appropriate
choice of gauge. Thus ${z}$ takes values on the manifold
$$\frac{U(N)}{U(N-1)\times U(1)},$$ which is better known as the
complex projective space $CP^{N-1}$. Alternatively we can carry the
gauge invariance along since it involves a compact gauge field and
hence only contributes a finite multiplicative factor in finite
volumes.

A similar limit yields the classical Sp$(N)$ problem with energy
\begin{equation}
E = -\frac{1}{N}\sum_{\langle ij \rangle} 
|\mathcal J_{\alpha\beta} z_{i\alpha}z_{j\beta}|^2.
\label{eq-E-classical-SpN}
\end{equation}
As advertised, in the following we will confine ourselves to the SU$(N)$ problem.


\section{Infinite $N$}
\label{section-infiniteN}

\subsection{General}
\label{section-infiniteN-general}

If we now take the limit $N\to\infty$, the partition function can be 
evaluated using the standard saddle-point approximation.  First, the constraint
$z^*_{i\alpha}z_{i\alpha} = N$ is enforced with the aid of a Lagrange
multiplier $\lambda_i$ on every site; the quartic interaction is
made quadratic at the expense of introducing an auxiliary complex
variable $Q_{ij}$ on every link:
\begin{equation}
Z = \int D\lambda\, DQ\, Dz\, e^{-\beta E[Q,\lambda,z]},
\end{equation}
where the effective energy is
\begin{eqnarray}
E[Q,\lambda,z] &=& 
-\sum_{\langle ij \rangle} 
(Q_{ij}z^*_{i\alpha} z_{j\alpha} +\mbox{C.c.} - N|Q_{ij}|^2)
+ i\sum_{i} \lambda_i (z^*_{i\alpha}z_{i\alpha}-N)
\nonumber
\\
&=& 
E[Q,\lambda,0] + \sum_{i,j}z^*_{i\alpha} {\mathcal H}_{ij} z_{j\alpha}. 
\label{eq-action-full}
\end{eqnarray}
Integration over the original variables $z_{i\alpha}$ yields 
a new effective energy
\begin{equation}
E[Q,\lambda] = N\left( 
\sum_{\langle ij \rangle} |Q_{ij}|^2
-i\sum_{i} \lambda_i + \beta^{-1} \ln{(\det{ {\mathcal H}[Q,\lambda]})}
\right).
\label{eq-action-Qlambda}
\end{equation}

In the limit $N\to\infty$ the dominant contribution to the integral of
$e^{-\beta E[Q,\lambda]}$ comes from the vicinity of a saddle
point,
\begin{equation}
\frac{\partial E[Q,\lambda]}{\partial Q_{ij}} = 0,
\hskip 5mm
\frac{\partial E[Q,\lambda]}{\partial \lambda_i} = 0.
\end{equation}
These conditions yield a set of self-consistent (mean-field) equations 
\begin{equation}
NQ_{ij} = \langle z_{i\alpha} z^*_{j\alpha} \rangle,
\hskip 5mm
N = \langle z^*_{i\alpha} z_{i\alpha} \rangle,
\label{eq-mf-eqs}
\end{equation}
where $\langle\ldots\rangle$ denotes averaging over a thermal ensemble 
with energy
\begin{equation}
\sum_{i,j}{\mathcal H}_{ij} z^*_{i\alpha} z_{j\alpha} 
= \sum_{i} \mu_i z^*_{i\alpha} z_{i\alpha}
-\sum_{\langle ij \rangle} 
(Q_{ij}z^*_{i\alpha} z_{j\alpha} +\mbox{C.c.}).
\label{eq-mf-energy}
\end{equation}
Thus $Q_{ij} = Q^*_{ji}$ can be thought of as a hopping amplitude and
$\mu_i = i\lambda_i > 0$ as a chemical potential. Solving these
equations requires either an explicit choice of gauge or a recognition
that any non-trivial solution will really be a set of gauge equivalent
solutions. Adding in all gauge equivalent saddle points would appear to
restore manifest gauge invariance but an important subtlety in this 
procedure is the subject of Section \ref{subsection-gauge}.

Fluctuations of $Q$ and $\lambda$ contribute an amount of order $1/N$
and can be neglected in the limit $N\to\infty$.  In this approximation, 
free energy per boson flavor is 
\begin{equation}
\frac{F[Q,\mu]}{N} = \sum_{\langle ij \rangle} |Q_{ij}|^2
-\sum_{i} \mu_i + T\ \mbox{Tr }{\ln{ {\mathcal H}[Q,\mu]}}.
\label{eq-mf-free-energy}
\end{equation}
The last term --- free energy of coupled harmonic oscillators $z_i$ --- 
involves the matrix ${\mathcal H}$ with the following elements:
\begin{equation}
{\mathcal H}_{ij} = 
\left\{
\begin{array}{rl}
\mu_i & \mbox {if $i=j$,} \\
-Q_{ij} & \mbox{if $i$ and $j$ are nearest neighbors,}\\
0 & \mbox{otherwise.}
\end{array}
\right.
\label{eq-mf-H}
\end{equation}
The saddle-point conditions (\ref{eq-mf-eqs}) become 
\begin{equation}
1 = \langle z_i z^*_i \rangle = T({\mathcal H}^{-1})_{ii},
\hskip 5mm
Q_{ij} = \langle z_i z^*_j \rangle = T({\mathcal H}^{-1})_{ij}.
\label{eq-mf-equip}
\end{equation}
(We have used the equipartition theorem for coupled harmonic oscillators 
$z_i$.)  

In what follows we will explore states preserving time reversal
symmetry; in such cases, one can choose a gauge where saddle-point
values $Q_{ij}$ are real.  The set of equations (\ref{eq-mf-H}) and
(\ref{eq-mf-equip}) defines mean-field solutions of the large-$N$
model.  This is a difficult nonlinear problem and an analytical
solution is not always possible.  However, there are a few helpful
general results that we spell out below.

\subsubsection{Solving for chemical potential}
\label{section-infiniteN-general-mu}

Any mean-field solution satisfies the equations
\begin{equation}
\mu_i = T + \sum_{j(i)} |Q_{ij}|^2,
\label{eq-mf-mu-via-Q}
\end{equation}
where the sum is taken over nearest neighbors of site $i$.  This 
result follows from the identity 
$\sum_{j}({\mathcal H}^{-1})_{ij} {\mathcal H}_{jk} = \delta_{ik}$
for $i=k$.

\subsubsection{Trivial solution $Q=0$}
\label{section-infiniteN-general-trivial}

Mean-field equations (\ref{eq-mf-H}) and (\ref{eq-mf-equip}) always
have a trivial solution with $Q_{ij}=0$ on all links and $\mu_i=T$.
($\mathcal H$ is proportional to the unit matrix: ${\mathcal H} 
= T \mathbbm 1$.)
As we will see, this solution becomes a local minimum of free energy
above the temperature $T_0 = 1$.  At high enough (but finite)
temperatures, this is the {\em global} minimum.

\subsubsection{Continuous phase transition at $T=1$}
\label{section-infiniteN-general-smallQ}

At a high enough temperature, the system is in the random phase with 
$Q_{ij}=0$ everywhere.  As temperature is lowered, we expect a 
transition into a phase where $Q_{ij}\neq 0$ on some links.  A 
continuous transition occurs when small fluctuations of $Q_{ij}$
have a vanishing cost in terms of free energy.  It therefore makes
sense to expand free energy in powers of $Q_{ij}$:  
\begin{eqnarray}
\frac{F[Q,\mu]-F[0,T]}{N} &=& \sum_{\langle ij \rangle} |Q_{ij}|^2
-\sum_{i} (\mu_i-T) + T\ \mbox{Tr }{\ln{ \frac{{\mathcal H}[Q,\mu]}{T}}}
\label{eq-mf-free-energy-diff}
\\ 
&=& \mbox{Tr }\left[\frac{{\mathcal Q}^2}{2} - {\mathcal M}+T
+ T \ln{\left(\frac{\mathcal M-Q}{T}\right)}\right]
\nonumber
\\
&=& \mbox{Tr }\left[
\frac{\mathcal Q^2}{2} 
- T \sum_{n=2}^\infty \frac{1}{n} 
\left(\frac{\mathcal Q - \mathcal M + T}{T}\right)^n
\right],
\label{eq-F-expansion}
\end{eqnarray}
where ${\mathcal M}$ and $-\mathcal Q$ are the diagonal and
off-diagonal parts of matrix $\mathcal H = \mathcal M - \mathcal Q$
(\ref{eq-mf-H}).  To eliminate the chemical potential $\mathcal M$, we
use the constraint $\langle z^*_i z_i \rangle = 1$, or $\partial
F/\partial \mathcal M_{ii} = 0$.  By varying free energy
(\ref{eq-F-expansion}) with respect to $\mathcal M$, we find that
$\mathcal M - T = \mathcal O(\mathcal Q^2)$.  Then, to order $\mathcal
Q^2$,
\begin{equation}
\frac{F[Q,\mu]-F[0,T]}{N} 
= \sum_{\langle ij \rangle} \frac{T-1}{T}|Q_{ij}|^2
+ \ldots
\label{eq-F-expansion-Q2}
\end{equation}

Eq. (\ref{eq-F-expansion-Q2}) shows that the random phase ($Q_{ij}=0$)
becomes {\em locally} unstable for $T<1$.  Therefore, {\em if} the
transition to a phase with $Q_{ij}\neq 0$ is continuous, it must occur
at the critical temperature $T_c=1$.

It turns out, however, that in many cases the transition is {\em
discontinuous} (see Table \ref{table-order} in Section
\ref{section-infiniteN-Landau}).  Therefore, we should not truncate
the expansion at order $Q^2$: a cubic term or an expressely negative
quartic one generally result in a first-order transition with $T_c>1$.
At intermediate temperatures $1<T<T_c$ the random phase $Q=0$ is
locally stable but is not a global minimum.

What determines the order of the transition?  Sokal and Starinets
\cite{SS} suggest that it is the number of dimensions.  They have
shown that cubic lattices with $d<3/2$ exhibit a continuous
transition, while those with $d>3/2$ have a discontinuous one.  In the
next few pages we will survey a few regular lattices with $d=0,1,2,3$
and $\infty$.  (The results are summarized at the beginning of Section
\ref{section-infiniteN-Landau} in Table \ref{table-order}.)  It will
be seen that the order of the transition, in fact, has nothing to do
with the number of dimensions!  The relevant concept is, instead, the
{\em local connectivity} of the lattice.  Typically, the existence of
short loops (of length 3 or 4) will make the transition discontinuous.

\subsection{Regular lattices composed of equivalent sites}
\label{section-infiniteN-Bravais}

\begin{figure}
\begin{center}
\epsfig{file=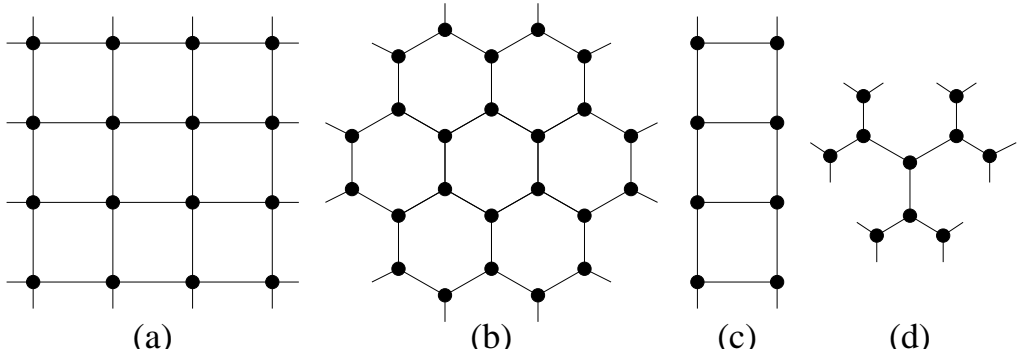,width=5.3in}
\end{center}
\caption{Examples of regular lattices composed of equivalent sites: 
(a) square, (b) honeycomb, (c) ladder, (d) Bethe lattice.}
\label{fig-lattices}
\end{figure}

To make further progress, consider lattices composed of equivalent
sites (Fig.~\ref{fig-lattices}).  We will investigate solutions in
which both the chemical potential and link variables are uniform, so
that $\mathcal H = \mu \mathbbm 1 - Q \mathcal A$, where $\mathcal A$
is the adjacency matrix ($\mathcal A_{ij}=A_{ji}=1$ for nearest
neighbors and 0 otherwise).  Note that the link strength $Q$ must be
real in order for the matrix $\mathcal H$ to be Hermitian.

The free energy now becomes a function of only two parameters $Q$ and $\mu$:
\begin{equation}
\frac{F(Q,\mu)-F(0,T)}{NV} = \frac{z Q^2}{2} - \mu + T 
+ \frac{T}{V} \sum_{k}\ln{\left(\frac{\mu-Q a_k}{T}\right)},
\label{eq-F-uniform}
\end{equation}
where $a_k$ are eigenvalues of the adjacency matrix $\mathcal A$ and $z$
is the coordination number of the lattice.  Variations with respect to 
$\mu$ and $Q$ yield the mean-field equations
\begin{eqnarray}
1 &=& \frac{T}{V} \sum_{k} \frac{1}{\mu - Q a_k},
\label{eq-mf-uniform-1}
\\
zQ &=& \frac{T}{V} \sum_{k} \frac{a_k}{\mu - Q a_k}
\label{eq-mf-uniform-zQ}
\end{eqnarray}
One can easily derive from these an analogue of Eq. (\ref{eq-mf-mu-via-Q}),
\begin{equation}
\mu - z Q^2 = T,
\label{eq-mf-mu-via-Q-uniform}
\end{equation}  
which can be used instead of Eq. (\ref{eq-mf-uniform-zQ}).

\subsubsection{$d=0$: two sites}
\label{section-infiniteN-Bravais-link}

The adjacency matrix has two eigenvalues, $\pm 1$, so that
Eqs. (\ref{eq-mf-uniform-1}-\ref{eq-mf-uniform-zQ}) are easily solved
to obtain an equation of state:
\begin{equation}
Q^2 (Q^2+T-1) = 0.  
\label{eq-of-state-link}
\end{equation}
At high temperatures, $T>1$, there is only a trivial solution $Q^2=0$. 
Below $T=1$, the global minimum of the free energy moves to $Q^2=1-T$
producing a continuous phase transition.  

\subsubsection{$d=0,1$: periodic chain}
\label{section-infiniteN-Bravais-loop}

\begin{figure}
\begin{center}
\epsfig{file=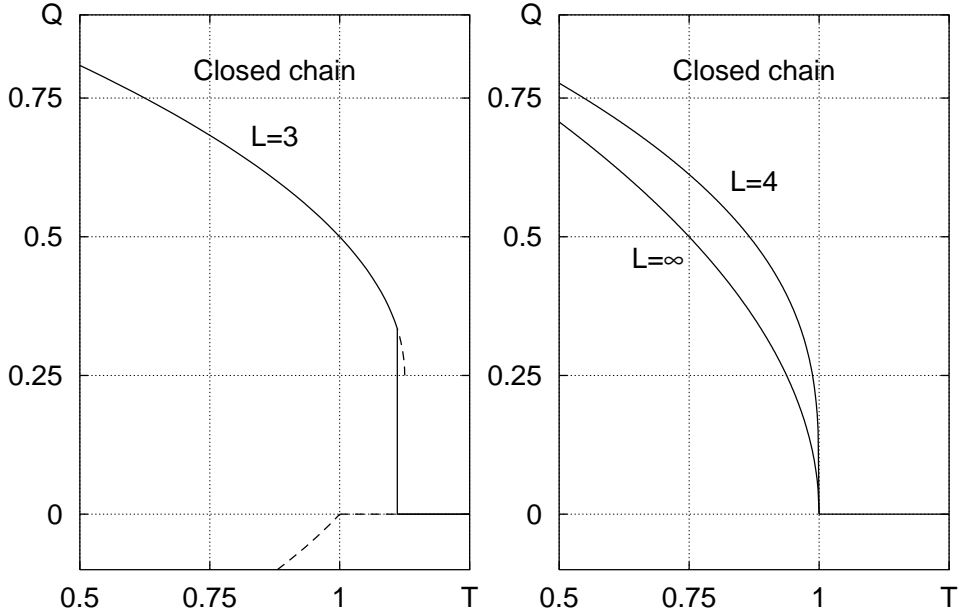,width=5in}
\end{center}
\caption{Left: First-order phase transition on a closed chain of 
length $L=3$.  Dashed lines indicate metastable states. 
Right: second-order transition on closed chains of length $L=4$
and $\infty$.
}
\label{fig-chain}
\end{figure}

For a chain of length $L$ with periodic boundary conditions, the
equation of state is obtained by solving Eqs. (\ref{eq-mf-uniform-1})
and (\ref{eq-mf-mu-via-Q-uniform}):
\begin{equation}
1 = \frac{1}{L}\sum_{n=1}^L \frac{T}{T-2Q\cos{(2\pi n/L)}+2Q^2}.
\end{equation}
In several cases this can be done analytically:
\begin{eqnarray}
L = 3: \hskip 5mm && Q^2 (T-1-Q+2Q^2) = 0, 
\label{eq-of-state-chain-3}
\\
L = 4: \hskip 5mm && Q^2 [T-1+(T-1+2Q^2)^2] = 0,
\label{eq-of-state-chain-4}
\\
L = \infty: \hskip 5mm && Q^2 (T-1+Q^2) = 0.
\label{eq-of-state-chain-inf}
\end{eqnarray}
These are shown in Fig. \ref{fig-chain}. 

\begin{figure}
\begin{center}
\epsfig{file=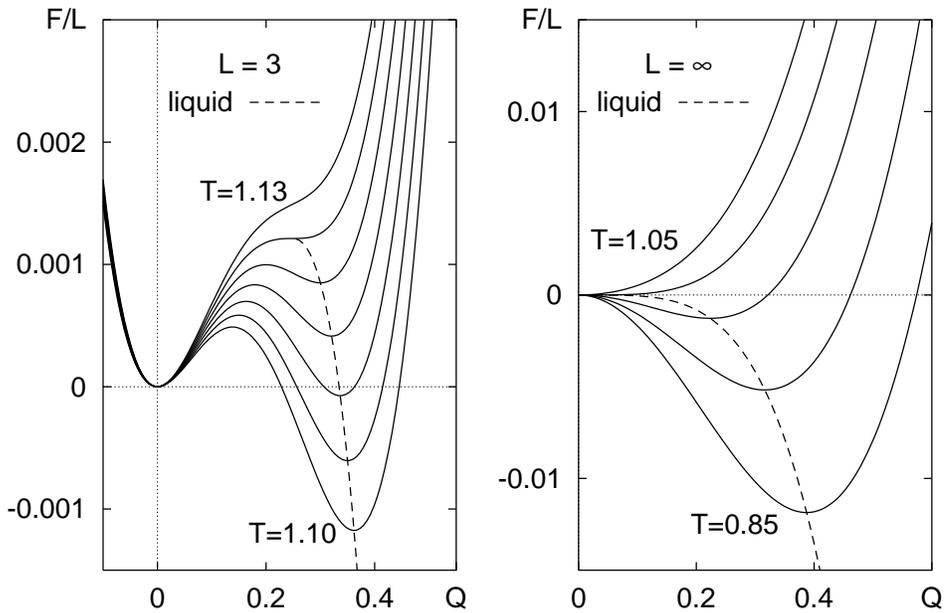,width=5in}
\end{center}
\caption{Free energy per site $F(Q,T)/L$, periodic chains with $L=3$
(left) and $L=\infty$ (right) for several temperatures around $T_c$.
Dashed lines trace minima of the free energy corresponding to the
liquid phase ($Q\neq 0$).  For $L=3$, there is a first-order transition
at $T_c=1.1107\ldots$ For $L=\infty$, the transition at $T_c=1$ is
continuous.}
\label{fig-F3}
\end{figure}

The equation for a 3-site chain acquires nontrivial solutions when $T
\leq 9/8$:
\[
Q_1 = 0, \hskip 5mm 
Q_2 = \frac{1-\sqrt{9-8T}}{4}, \hskip 5mm
Q_3 = \frac{1+\sqrt{9-8T}}{4}.
\]
There is a first-order transition at $T=T_c = 1.1107\ldots$ At that
temperature, the absolute minimum of the free energy $F(Q,T)$ jumps
from $Q_1=0$ to $Q_3>1/4$, see Fig. \ref{fig-F3}.  

At $L = 4$, the transition is continuous, although $Q$ rises unusually 
steeply for a mean-field theory: $Q \sim 2^{-1/2}(T_c-T)^{1/4}$.  
For all $L>4$, including the infinite chain, the transition is continuous
with the usual exponent $\beta=1/2$.  

\subsubsection{$d=2$: square lattice}
\label{section-infiniteN-Bravais-square}

The equation of state on the infinite square lattice is again obtained by
using the Eqs. (\ref{eq-mf-uniform-1}) 
and (\ref{eq-mf-mu-via-Q-uniform}): 
\begin{equation}
1 = \int_0^{2\pi} \frac{{\rm d} k_1}{2\pi}
\int_0^{2\pi} \frac{{\rm d} k_2}{2\pi}\ 
\frac{T}{T-2Q(\cos{k_1}+\cos{k_2})+4Q^2}.
\label{eq-of-state-square}
\end{equation}
There is only a trivial solution $Q=0$ at high temperatures, $T \gg
1$; in the opposite limit, $T \ll 1$, there are two solutions, $Q_1=0$
and $Q_2 \approx 1$, the latter being the global minimum of the free
energy.  The transition between the phases with $Q=0$ and $Q\neq 0$ is
discontinuous and takes place at $T_c>1$.  For if the phase transition
were continuous, it would occur at $T=1$, as we have argued
previously; $Q=0$ would still be the global minimum of the free energy
at the critical temperature.  Instead, we will see that at $T=1$ the
system is already in the phase with $Q\neq 0$, so that $T_c>1$.

To see that Eq. (\ref{eq-of-state-square}) has more than one solution
at $T=1$, look at the behavior of its right-hand side.  It diverges
logarithmically at $Q=1/2$ and tends to 0 as $Q\to\infty$.  By
continuity, there must be a solution for $Q>1/2$ --- in addition to
the trivial one, $Q=0$.
Because at $T=1$ the system is already in the phase with $Q\neq0$, the phase
transition takes place at a higher temperature and is discontinuous.

\begin{figure}
\epsfig{file=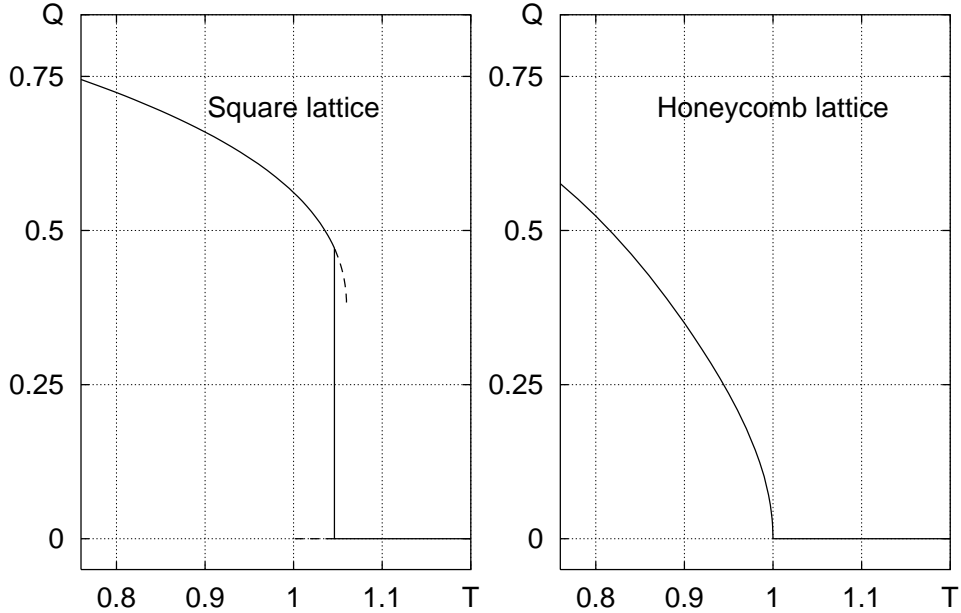,width=5in}
\caption{Left: First-order phase transition on the square lattice.
Dashed lines indicate metastable states. Right: second-order transition
on the honeycomb lattice.
}
\label{fig-square-honey}
\end{figure}

The equation of state (\ref{eq-of-state-square}) can be integrated
to obtain a closed form, 
\begin{equation}
1 + \frac{4Q^2}{T} = \frac{2}{\pi}\ {\bf K}\!\left(\frac{4Q}{T+4Q^2}\right),
\end{equation}
where ${\bf K}(k)$ is the complete elliptic integral of the first
kind.  The dependence $Q(T)$ is shown in Fig. \ref{fig-square-honey}. 

\subsubsection{$d=2$: honeycomb lattice}
\label{section-infiniteN-Bravais-honeycomb}

One might think that the order of the transition is determined by 
the dimensionality of the lattice.  This is not the case.  While the 
transition on the two-dimensional square lattice is first-order, 
its counterpart on the honeycomb lattice is continuous 
(Fig. \ref{fig-square-honey}). 

\subsubsection{$d=3$: cubic lattice}
\label{section-infiniteN-Bravais-cubic}

This case has been discussed previously by Sokal and Starinets
\cite{SS} for the SU$(N)$ case and by DeSilva {\em et al.} \cite{DMZ}
for Sp$(N)$.  There is a first-order transition at $T_c>1$.  Unlike in
lower dimensions, below $T_c$ the system is in a ferromagnetic state
that breaks the SU($N$) symmetry.  In an appropriate basis,
\[
\langle z^*_{i1} z_{i1} \rangle \neq \langle z^*_{i2} z_{i2} \rangle
= \langle z^*_{i3} z_{i3} \rangle 
= \ldots = \langle z^*_{iN} z_{iN} \rangle. 
\] 

It is interesting to note that the ferromagnetic transition in the
$d=3$ Heisenberg model --- the SU(2) case --- is continuous.  The
change to a first-order transition at large $N$ has a well-documented
analogue in the two-dimensional Potts model \cite{Potts}.  There, the
ferromagnetic transition is continuous for $q \leq 4$ states and
discontinuous for $q>4$. We will see below why the Potts model
is a simpler analog of our problem.

\subsubsection{$d=\infty$: Bethe lattice}
\label{section-infiniteN-Bravais-Bethe}

The Bethe lattice is another regular structure with
equivalent sites (Fig. \ref{fig-lattices}).  When the number of
nearest neighbors $z \geq 3$, the number of $n$-th neighbors grows
faster than any power of $n$.  In this sense, the Bethe lattice has an
infinite number of dimensions.

The spectrum of the adjacency matrix for the Bethe lattice is well known
\cite{Kesten}.  Its eigenvalues fill the interval $|a| < 2\sqrt{z-1}$
with density 
\begin{equation}
\rho(a) = \frac{z \sqrt{4(z-1)-a^2}}{2\pi (z^2-a^2)}.
\label{eq-rho-Cayley}
\end{equation}
Eq. (\ref{eq-mf-uniform-1}) can now be integrated.  The equation of
state is
\begin{equation}
Q^2 (Q^2+T-1) = 0.  
\label{eq-of-state-Bethe}
\end{equation}
Note that it is independent of the coordination number $z$ and
is therefore the same as that of an infinite chain (the Bethe lattice
with $z=2$).  The phase transition is continuous.

\subsection{Expansion of the Landau free energy}
\label{section-infiniteN-Landau}

\begin{table}
\caption{Order of the $N\to\infty$ phase transition and the exponent $\beta$ 
for regular lattices.}
\label{table-order}
\begin{center}
\begin{tabular}{lcccc}
\hline
\hline
lattice & $d$ & order of transition & $\beta$ & shortest cycle\\
\hline
chain, $L=3$ & 0 & 1st & 0 &   3  \\
ladder  & 1 & 1st & 0 &   4  \\
triangular&2& 1st & 0&  3   \\
square  & 2 & 1st & 0 &   4  \\
cubic   & 3 & 1st & 0 &   4  \\
\hline
chain, $L=4$ & 0 & 2nd & 1/4 &   4  \\
\hline
2 sites & 0 & 2nd & 1/2 &  none\\
chain, $L=\infty$   & 1 & 2nd & 1/2 &  none\\
honeycomb&2 & 2nd & 1/2 &   6  \\
diamond & 3 & 2nd? & 1/2? &   6  \\
Bethe   & $\infty$ & 2nd & 1/2 & none\\
\hline
\hline
\end{tabular}
\end{center}
\end{table}

Our survey of regular lattices is summarized in Table
\ref{table-order}.  It demonstrates that the order of the phase
transition has nothing to do with the dimensionality of the system.
This is not surprising: in the continuous version of the transition,
the correlation length vanishes, instead of becoming infinite.
Therefore, long-distance properties, such as dimensionality, are not
important (in the limit $N\to\infty$).  Rather, details of the
transition are determined by the local structure of the lattice.  This
connection is evident in Table \ref{table-order}: the transition is
discontinuous when the lattice has closed loops of length 3 or 4.
(The 4-site chain is a border case: although the transition is
continuous, the exponent $\beta=1/4$ is unusually small.)

To substantiate this claim, we expand the free energy
(\ref{eq-F-uniform}) in powers of $Q$:  
\begin{equation}
\frac{F(Q,\mu)-F(0,T)}{NV} = \frac{z Q^2}{2} 
- \frac{T}{V} \sum_{k}\sum_{n=2}^\infty \frac{1}{n} 
\left(\frac{Q a_k-\mu+T}{T}\right)^n.
\label{eq-F-uniform-expansion1}
\end{equation}
The term $\mathcal O(Q^2)$ was written out in
Eq. (\ref{eq-F-expansion-Q2}).  To obtain an expansion to order
$\mathcal O(Q^4)$, $\mu$ should be evaluated --- to order $Q^2$
--- using the constraint $\partial F/\partial \mu=0$, Eq. 
(\ref{eq-mf-uniform-1}).
We then obtain 
\begin{equation}
\frac{F(Q,\mu) - F(0,T)}{NV} = \frac{T-1}{T}\, \frac{z_2 Q^2}{2}
- \frac{z_3 Q^3}{3T^2} + \frac{(2z_2^2 - z_4)Q^4}{4T^3}
+ \mathcal O(Q^5).
\label{eq-F-order-4}
\end{equation}
The integers
\[
z_n = \frac{1}{V} \sum_{k} a_k^n = \frac{1}{V} {\rm Tr} \mathcal A^n
\]
give the number of distinct loops of length $n$ starting from some
lattice node.  In particular, $z_2$ is simply the number of nearest
neighbors $z$.  

At high temperatures, the minimum of free energy (\ref{eq-F-order-4})
is at $Q=0$.  For $T>1$, this point is a local minimum of $F$, at
$T<1$ it becomes a local maximum.  A necessary condition for a
continuous phase transition at $T_c=1$ is the absence of a cubic term in
this expansion \cite{LL}.  A negative quartic term will also make the
transition discontinuous.  

\subsubsection{Cubic term}

A cubic term is only possible on lattices containing loops of length
3, in which case $z_3\neq0$.  Therefore the transition is
discontinuous in a closed chain of length 3 and on the triangular
lattice.

\subsubsection{Quartic term}

The sign of the quartic term is determined by the presence of cycles 
of length 4.  Generally, if there are no such loops, the quartic term 
is positive and the transition is likely continuous (barring a cubic
term or some pathological behavior in higher orders).  All lattices
listed in the lower part of Table \ref{table-order} look like trees
if explored at depths up to 4.  Continuous phase transitions are then
possible.  

The presence of loops of perimeter 4 makes the quartic term negative
or --- in special cases --- zero.  This alters the character of the
transition and makes it first order.  The remaining lattices in Table
\ref{table-order} (4-site chain, ladder, square, cubic) all have loops
of length 4.  

\begin{figure}
\begin{center}
\epsfig{file=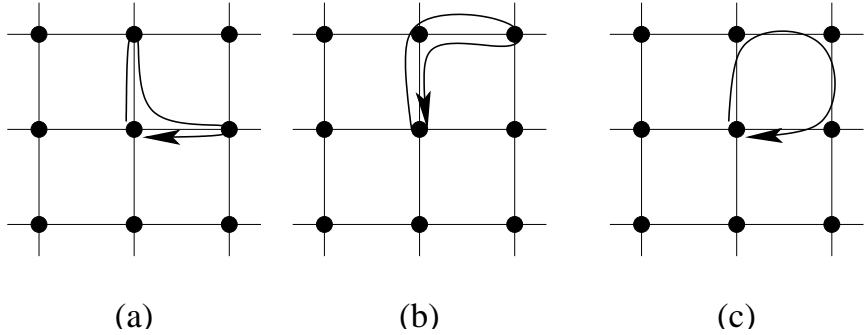,width=4.5in}
\end{center}
\caption{
There are $z^2 + z(z-1)$ noncyclic closed paths (a) and (b) of length 4
--- in addition to
cycles (c).
}
\label{fig-paths}
\end{figure}

To prove these conjectures, let us evaluate the prefactor of the
quartic term, $2z_2^2-z_4$, that is responsible for its sign.  On a
tree without loops, the number of round-trip paths of length 4 is
$z^2+z(z-1)$, see Fig. \ref{fig-paths} (a,b).  In this case,
$2z_2^2-z_4 = z>0$ and the quartic term is positive.

If, in addition, $z_4$ contains $l_4$ loops [Fig. \ref{fig-paths}
(c)], the quartic term is proportional to $z-l_4$.  It is easy to see
that the number of the loops $l_4$ cannot be smaller than $z$ (if they
are present at all), hence the quartic term cannot be positive when
such loops are present.  

The quartic term is expressely negative for a ladder ($z=3$, $l_4=4$),
a square lattice ($z=4$, $l_4=8$), and a cubic lattice ($z=6$,
$l_4=24$), hence first-order transitions.  A chain of length 4 ($z=2$,
$l_4=2$) is a special case: the quartic term just vanishes.  The
transition is still continuous --- thanks to a positive term $\mathcal
O(Q^6)$? --- but $Q$ rises more steeply than $(T_c-T)^{1/2}$,
Fig. \ref{fig-chain}.

\subsection{It's the connectivity, stupid!}

In this section, we have investigated a phase transition that occurs in an
SU($N$) ferromagnet at infinite $N$.  It is a transition between a 
seriously disordered high-$T$ phase with no correlations between spins
whatsoever and a low-$T$ phase that is either a paramagnet with
short-range correlations (in low dimensions $d \leq 3$), or possibly
a ferromagnet.  

The transition can be first or second-order, as noted by several
authors previously \cite{SS,DMZ}.  Here we have uncovered what
determines the order of the transition: it is the local connectivity
of the lattice, rather than its dimensionality.  In hindsight, this is
not surprising because --- in its continuous version --- this is a
transition between phases with finite and {\em zero} correlation
lengths, so it should be more sensitive to local features (presence of
loops) than to long-distance ones (number of dimensions).  

We have substantiated our claim by expanding the free energy in powers
of an order parameter $Q_{ij} = \langle z_i z_j^\dagger\rangle/N$
living on links.  The presence of loops of length 3 leads to the
existence of a cubic term in this expansion; loops of length 4 tend to
make a quartic term negative.  In either case, by the standard Landau
argument, the transition becomes discontinuous. Finally a caveat on
connectivity: our statements are for the purely nearest neighbor model.
Interactions of further range will effectively change the connectivity.


\section{Large $N$}
\label{section-finiteN}

So far we have discussed the limit of infinite $N$, when the
saddle-point approximation is exact.  We turn next to what happens when 
$N$ is finite.  One aspect is clear: the high-temperature phase is no 
longer perfectly disordered.  Nevertheless, 
for sufficiently large $N$, the mean-field theory should be 
a good starting point for the analysis. To proceed further we will ask
three questions. First, whether the $N=\infty$ transition is characterized
by an ordering that can also characterize a phase transition at
finite $N$. Second, whether the finite $N$ phase transition can be 
characterized by a different ordering that still  continuously connects 
to the $N=\infty$ transition. Third, we will examine the nature of
the finite $N$ corrections to see what they suggest about the fate of the 
transition.

We will consider these questions in turn, starting with the first question 
which has been discussed previously under the rubric of Elitzur's theorem 
for gauge theories.

\subsection{Spontaneous breaking of a gauge symmetry at $N=\infty$}
\label{subsection-gauge}

As noted at the outset, our class of problems exhibit a local gauge
invariance. In our $N=\infty$ analysis we picked a gauge to identify a
saddle point and left open the option of not picking a gauge and
simply adding in all the gauge equivalent saddle points. While this
would appear to be a manifestly gauge invariant way of proceeding, it
hides an interesting anomaly noted most recently by Sokal and
Starinets \cite{SS}: the local U(1) gauge symmetry appears to be
spontaneously broken in the low-temperature phase if one invokes the
usual criterion of a response to an infinitesimal field taken to zero
post-thermodynamic limit.\footnote{Spontaneous breaking of a gauge
symmetry at large $N$ has been investigated by several authors
\cite{Samuel,CG}}. So this answers our first question: there {\it is}
a broken symmetry at $N=\infty$ but it cannot be broken at any finite
$N$ by virtue of Elitzur's theorem \cite{Elitzur}.  Sokal and
Starinets go further and use this observation to conclude that ``it
seems unlikely that such a transition can survive to finite $N$.'' We
show in the following, by a simpler example, that this is a red
herring and that at least in the case of a first order transition
there is no contradiction between the restoration of Elitzur's theorem
and the survival of the transition.  The example involves the Potts
model, which we will rewrite as an Ising gauge theory. This will
enable us to appeal to well known results on the Potts model to
establish the phase diagram of the theory.  Altogether we will see the
breakdown of gauge invariance is a consequence of something
special---the divergence between the exchange constant and the
transition temperature in the infinite $N$ limit.

First let us note the violation of Elitzur's theorem explicitly.
The gauge symmetry in question is evident from the form of energy
(\ref{eq-E-classical}).  Changing the phase of the oscillator variable
$z_{i\alpha} \mapsto z_{i\alpha} e^{{\rm i}\chi_i}$ leaves the energy
invariant.  However, the link variable $Q_{ij} = \langle z_{i\alpha}
z^*_{j\alpha}\rangle/N$ is certainly not invariant: $Q_{ij} \mapsto
Q_{ij}e^{{\rm i}(\chi_i-\chi_j)}$.  Thus the absolute phases of
$Q_{ij}$ are immaterial.  In particular, changing the sign of all link
variables (on a bipartite lattice!) gives a physically equivalent
configuration.  Nevertheless, it can be shown that adding an
infinitesimal gauge-fixing term to the energy,
\begin{equation}
E(\eta) = -\sum_{\langle ij \rangle}
\left[\frac{|z^*_{i\alpha}z_{j\alpha}|^2}{N} + \eta
(z^*_{i\alpha}z_{j\alpha} + 
z^*_{j\alpha}z_{i\alpha})\right],
\label{eq-E-eta}
\end{equation}
leads to a non-analytic behavior of the free energy near $\eta \to 0$:
\begin{equation}
\frac{F[Q,\mu,\eta]}{N} = \frac{F[Q,\mu,0]}{N} 
- 2|\eta| \sum_{\langle ij \rangle}  |Q_{ij}|
+ \mathcal O(\eta^2).
\label{eq-F-eta}
\end{equation}
(The limit $N\to\infty$ should be taken first to ensure validity of
the mean-field treatment.)  The nonalyticity of the free energy
(\ref{eq-F-eta}) means that the values of $Q_{ij}$ are frozen either
near $+|Q_{ij}|$ or $-|Q_{ij}|$, depending on the sign of $\eta$.
Thus the local U(1) gauge symmetry is spontaneously broken in a phase
with $|Q_{ij}|\neq 0$.

\subsubsection{Question}
\label{section-gauge-question}

Why is a gauge symmetry spontaneously broken at large $N$?  What
happens when the number of flavors is large but finite?  To answer
these questions, one could study thermodynamics of an SU($N$)
ferromagnet in the presence of a small symmetry-breaking term
(\ref{eq-E-eta}).  The partition function of an infinite chain (or 
just of a single link) can be evaluated for any finite $N$ using
the transfer matrix: 
\begin{equation}
[Z(\beta,\eta)]^{1/L} = 2(N-1) \int_0^{\pi/2} \sin^{2N-3}{\theta}\,
\cos{\theta}\, d\theta \
{\rm e}^{N\beta(\cos^2{\theta}-\eta\cos{\theta})}
\label{eq-Z-eta}
\end{equation}
(see the companion paper \cite{FT} for details on the integration
measure).

Evaluation of the partition function (\ref{eq-Z-eta}) is a feasible,
though not particularly straightforward task.  To keep technical
details to a minimum, we have chosen to study a similar phenomenon 
in the Potts model.  Because it involves discrete degrees of freedom, 
the broken gauge symmetry is also discrete ($Z_2$).  

\subsubsection{Insights from the Potts model}
\label{section-gauge-Potts}

We define the Potts model \cite{Potts} in terms of unit vectors ${\bf
S}_i$ that can point along orthogonal axes in a $q$-dimensional
internal space: the energy is
\begin{equation}
E = -\sum_{\langle ij \rangle} ({\bf S}_i \cdot {\bf S}_j)^2,
\hskip 5mm 
{\bf S}_i = \pm\hat{\bf e}_1,\ \pm\hat{\bf e}_2, \ldots,\ \pm\hat{\bf e}_q,
\hskip 5 mm
\hat{\bf e}_{m}\cdot \hat{\bf e}_n = \delta_{mn}. 
\label{eq-Potts-Z2}
\end{equation}  
We have doubled the number of states per site compared to the usual
amount.  As a result, the energy is invariant under a local $Z_2$
symmetry ${\bf S}_i \mapsto -{\bf S}_i$.  Apart from a multiplicative
factor, the partition function is identical to that of the Potts
model.  Parametrization (\ref{eq-Potts-Z2}) is a discrete analogue of
representing SU($N$) spins in terms of Schwinger bosons
(\ref{eq-E-classical}).

We can construct an infinite $q$ treatment along the same lines as
the large $N$ treatment discussed previously: we decouple the
quartic interaction in favour of a gauge field:
 \begin{equation}
 {\rm e}^{\beta ({\bf S}_i \cdot {\bf S}_j)^2 }
 = \sqrt{\beta/\pi} \int_{-\infty}^{\infty}\! {\rm d}Q_{ij} \ 
 {\rm e}^{-\beta (Q_{ij}^2 - 2 Q_{ij}{\bf S}_i \cdot {\bf S}_j)}.
 \end{equation}
and then integrate the spins out.
The result of this analysis is the prediction of
a finite temperature first order phase transition from a phase with zero
correlation length as the temperature is lowered. 
Again the infinite $q$ system breaks the Ising gauge invariance.

Now we know that for any finite $q$, the Potts model has a phase 
transition in $d=2$ or more but not in $d=1$ and that the transition
is indeed first order at large $q$ in $d \ge 2$ dimensions.\footnote{This 
treatment correctly predicts the order of the transition.  However, the
low-temperature phase is not a paramagnet: in $d \ge 2$ dimensions, the
discrete global symmetry of the Potts model can be spontaneously broken.
In order to characterize a ferromagnetic phase, one must introduce an
appropriate order parameter.  This complication is absent in the SU$(N)$ 
model in $d=2$ in view of the Mermin-Wagner theorem.} We also know that
the finite $q$ problem cannot break gauge invariance. Evidently the
prediction of a first order transition in a dimension where it has
every reason to be robust, is {\it not} vitiated by the breaking
of local gauge invariance.

Indeed one can use this example to see where the breaking comes from.
Consider for simplicity the one dimensional chain.
The free energy density of an infinite chain 
can be evaluated with the aid of the transfer matrix yielding
\begin{equation}
f(\beta) = 
\lim_{L\to\infty} F(\beta)/L = -\beta^{-1} \ln{({\rm e}^\beta+q-1)}. 
\end{equation}
This is an analytic function of $\beta$ for any finite $q$.  At large
$q$, 
\begin{equation}
f(\beta) \sim 
\left\{
\begin{array}{ll}
-\beta^{-1} \ln{q} & \mbox{ if } \beta < \ln{q}, \\
-1 & \mbox{ if } \beta > \ln{q}.
\end{array}
\right.
\end{equation}
In the limit $q\to\infty$, the free energy develops a kink at $\beta =
\ln{q}$, so that there is a first-order phase transition.

The $Z_2$ gauge symmetry is broken in the low-temperature phase
($\beta > \ln{q}$).  To see this, add a symmetry-breaking term
$-\eta\, {\bf S}_i \cdot {\bf S}_j$ to the energy of every bond and
evaluate the free energy in its presence:
\begin{equation}
f(\beta,\eta) 
= -\beta^{-1} \ln{({\rm e}^\beta \cosh{\beta \eta} +q-1)}. 
\end{equation}
The expectation value of the gauge-dependent quantity 
\begin{equation}
\langle {\bf S}_i \cdot {\bf S}_{i+1} \rangle
= -\frac{\partial f(\beta,\eta)}{\partial \eta}
= \frac{\sinh{\beta \eta}}{\cosh{\beta \eta} +(q-1){\rm e}^{-\beta}}
\end{equation}
(the analogue of $\langle z^\dagger_i z_i\rangle$) depends on the
order of limits $\eta\to 0$ and $q\to\infty$:
\begin{equation}
\langle {\bf S}_i \cdot {\bf S}_{i+1} \rangle \sim 
\left\{
\begin{array}{ll}
\beta \eta & \mbox{ if }\eta \ll 1/\ln{q} \ll 1, \\
{\rm sgn}\, \eta & \mbox{ if } 1/\ln{q} \ll \eta \ll 1, 
\end{array}
\right.
\label{eq-order-of-lim}
\end{equation}
If the limit $\eta\to 0$ is taken first (at finite $q$), the gauge
symmetry remains intact.  Reversing the order of limits ($q\to\infty$
first) leads to spontaneous breaking of the gauge symmetry.  

\subsubsection{Answer}
\label{section-gauge-answer}

What have we learned from the Potts model?  Gauge symmetry can be
spoiled by adding a term that prefers one gauge configuration over all
others.  The symmetry is spontaneously broken if gauge selection
reliably occurs even when the gauge-fixing term is small.  The example
we have just worked out shows that the there may be different degrees
of smallness: the symmetry-breaking perturbation can be compared to
the interaction strength, as well as to temperature.  The presence of
a large number of flavors in the model pushes the temperature scale
down (as $1/N$ in the Schwinger-boson case or as $1/\ln{q}$ in the
Potts model).  Thus a nominally small gauge-fixing term ($\eta \ll 1$)
can still be large enough ($T \ll \eta \ll 1$) to pick out a gauge.
Most importantly, this does not rule out a persistence of the
transition when $q$ is finite, contradicting the belief of Sokal and
Starinets.


\subsection{Gauge-invariant order parameters?}

The next order of business is to ask whether a gauge-invariant
order parameter can discriminate between the putative finite
$N$ versions of the two phases. Note that both are $SU(N)$
invariant. In our discussion of $d \ge 3$ later we will briefly
consider the new features that arise when that can be broken as
well.

Away from infinite $N$, $\langle Q_{ij} \rangle =0$ by gauge
invariance while the amplitude $\langle |Q_{ij}| \rangle$ will
be non-zero in both phases although it could jump across the
transition. At any rate it cannot serve as an order parameter.
This leaves the gauge invariant fluxes that remain when the
fluctuations of the amplitude are integrated out. In $d=1$
this sector is empty. In higher dimensions these would be 
described, at large $N$, by a weakly coupled $U(1)$ gauge theory.
While the coupling may jump across the phase boundary, at
sufficiently large $N$ is will be small on either side and
we expect that the gauge fluctuations will be in the same phase
on either side, i.e. barely confining in $d \le 3$ while deconfining
in $d \ge 4$. In any event we do not expect a qualitative
distinction to develop between the two phases and hence the
transition between them will have the character of a liquid-gas
transition, and not an order-disorder transition. For this
reason, we expect the transition to be generically first order
if it exists. A qualitative distinction {\it could} develop at 
smaller $N$ in $d \ge 3$ with a change in the character of
the transition. We will briefly return to this later in
our comments on $d \ge 3$.

An important comment is in order. From the perspective of the original
quantum magnetic problem, we are operating at high temperatures and
large ``spins''. In this limit the temporal dimension has shrunk to 
zero and so we are computing Wilson loops which are not really
diagnostic of the free energy cost of separating two spinons.
The latter requires a  computation of the temporal Polyakov
loops and shows deconfinement consistent with the notion that
the paramagnet is insensitive to the insertion or removal of
a local fixed spin. 

\subsection{No phase transition in $d \leq 1$}

Finite lattices and lattices in $d=1$ dimensions simply cannot have a 
phase transition when interactions are short ranged and $N$ is finite.
An infinite $N$, in essence, provides an extra dimension with a
long-range interaction [every flavor interacts with every other one,
see Eq. (\ref{eq-H-quantum1})] which allows this conclusion to be
evaded.  Thus in $d \leq 1$ the transition will smooth into a crossover 
near $T=1$.  An explicit solution of the one-dimensional SU$(N)$ and
Sp$(N)$ chains in the companion paper by Fendley and Tchernyshyov
\cite{FT}
confirms this; we direct the reader there for details of their method.

Here we content ourselves with following the smearing of the infinite-$N$ 
transition into a crossover in the case of a single link.  The effective free energy 
in this case is 
\begin{equation}
F(Q, {\rm i}\lambda_1, {\rm i}\lambda_2)
= N[|Q|^2 - {\rm i}(\lambda_1+\lambda_2)N 
+ \beta^{-1} \ln{(-\lambda_1 \lambda_2 -|Q|^2)}.]
\end{equation}
As a function of $\lambda_1$ and $\lambda_2$, it has a saddle point for 
real and positive 
\[
{\rm i}\lambda_1 = {\rm i}\lambda_2 = \mu(Q,T)
= T/2 + \sqrt{(T/2)^2 + |Q|^2}.
\]  
Integrating out Gaussian fluctuations of $\lambda$ gives a nonsingular
contribution of order 1 to the free energy of the link:
\[
F(Q) = N \left\{
|Q|^2 - 2\mu(Q,T) + T \ln{[\beta \mu(Q,T)]}
\right\} + \mathcal O(1).
\]
This free energy has a minimum at $Q=0$ if $T>1$; otherwise, the minimum
is at $|Q|^2 = 1-T + \mathcal O(1/N)$.  For large $N$ and $T$ near 1, the
partition function evaluates to 
\[
Z(\beta) = \int\! {\rm d}Q^* {\rm d}Q \ {\rm e}^{-\beta F(Q)}
\sim {\rm e}^{N(1+\beta-\ln{\beta})} 
\int_{T-1}^\infty \! {\rm d}x \ {\rm e}^{-Nx^2/2}.
\]
As temperature crosses 1, energy of the link smoothly changes from 
being $\mathcal O(N)$ to being $\mathcal O(1)$:
\[
E(T) = -\frac{{\rm d}\, \ln{Z(\beta)}}{{\rm d}\beta} 
\sim \left\{
\begin{array}{lll}
N(T-1) & \mbox{ if } T<1 & \mbox{ and } 1/\sqrt{N} \ll |T-1| \ll 1,\\
-\sqrt{\pi N/2} & & \mbox{ if } |T-1| \ll 1/\sqrt{N},\\
1/(1-T) & \mbox{ if } T>1 & \mbox{ and } 1/\sqrt{N} \ll |T-1| \ll 1.
\end{array}
\right.
\]
The width of the crossover is $\mathcal O(N^{-1/2})$.  As $N\to\infty$, 
the crossover turns into a phase transition.

\subsection{Liquid-gas transition in $d=2$}

This is the most interesting case for our purposes. For one thing, the
application of large $N$ methods has been most influential with respect
to ground state properties in $d=2$. For another, the finite temperature
problem cannot break the SU$(N)$ symmetry in $d=2$ by Mermin-Wagner
and so the intra-paramagnetic liquid-gas transition we have been considering 
is the only possibility for a thermal transition. 

\subsubsection{First-order cases}

In these cases, e.g. the square and triangular lattices,
the transition should survive at sufficiently large
$N$. Briefly, the infinite $N$ transition involves the crossing
of two separated saddle points. At large $N$ we expect corrections
to the contributions from the saddle points which are subdominant
in $N$ but these should not affect the existence of a crossing. The
only way this can go wrong is if a whole host of other saddle points
enter the finite $N$ computation and their entropy overcomes the
energy cost. This is what happens in $d=1$ where the other saddle
points are domains walls (instantons) that turn the first order 
transition into a cross-over. (One can produce a first order transition
in the infinte-$N$ problem in $d=1$ by adding a second neighbor coupling.)

In $d=2$, as long as there is a finite surface tension per flavor
between the two phases at $N=\infty$, any domain walls of order the
system size should be exponentially suppressed and hence the phase
transition should survive at sufficiently large $N$. As one of the two
phases has zero correlation length, we have carried out this
computation to see if the magnitude of the surface tension is
anomalously small.

\begin{figure}
\begin{center}
\epsfig{file=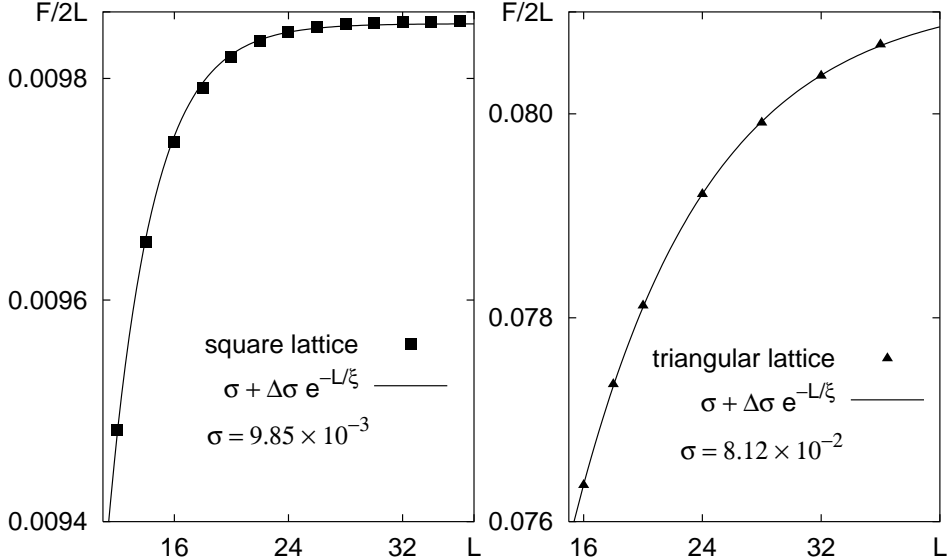,width=5in}
\end{center}
\caption{Free energy per unit length of a domain wall on square and
triangular lattices at their transition temperatures.  Lines are fits
to $F/2L = \sigma + \Delta\sigma\, {\rm e}^{-L/\xi}$, where $\sigma$
is the interface tension for an infinite domain wall, $L$ is the
system size, and $\xi$ is the correlation length of Schwinger bosons
in the liquid phase.}
\label{fig-tension}
\end{figure}

The computation is carried out by solving numerically the set of
mean-field equations (\ref{eq-mf-H}) and (\ref{eq-mf-equip}) on an $L
\times L$ square or triangular lattice at the coexistence temperature
of an infinite system.  We have used periodic boundary conditions.  A
starting configuration includes two domains (liquid and gas) separated
by two domain walls of total length $2L$.  The free energy per unit
length of the domain wall is fit to a simple form
\[
F/2L = \sigma + \Delta\sigma\, {\rm e}^{-L/\xi}.
\]
The second term, a finite-size effect, contains the spinon correlation
length $\xi$ determined in the uniform liquid state leaving only two 
fitting parameters, interface tension $\sigma$ and $\Delta\sigma$
(Fig.~\ref{fig-tension}).  

For the square lattice this procedure yields a surface tension $0.01$
per flavor per lattice constant and an interface of width $\xi \approx
3$ lattice constants.  For the triangular lattice this yields a
surface tension $0.08$ per flavor per lattice constant and an
interface of width $\xi\approx 9$ lattice constants. The surface
tension is higher on a triangular lattice as expected since the first
order transition is stronger in this case at $N=\infty$.

As the surface tension is finite, we conclude that immediately
away from $N=\infty$ both phases will get dressed by local fluctuations,
whose detailed theory at large $N$ is beyond our means at this point,
but the transition will survive. We
expect that the resulting line of first order transitions in the
$N,T$ plane will terminate, as in the liquid-gas problem, in a
critical end point governed by the $d=2$ Ising critical 
theory.\footnote{Similarly to density in the liquid-gas problem,
spin correlations $|Q_{ij}|^2$ are discontinuous across the transition
and thus play the role of an ``order parameter''.}  
Of course,
the location of this critical end point will be non-universal and
is unlikely to lie at an integer value of $N$.
In all of this, in a rough sense, valid near $N=\infty$, the temperature 
(or, rather, $T-T_c$) plays the role of the field that discriminates 
between the two phases, by favoring the gas or liquid saddle point; while
$N$ itself is an overall multiplicative factor in the free energy 
(\ref{eq-mf-free-energy-diff}), and therefore it acts as 
the inverse temperature. Indeed, infinite $N$ corresponds to zero temperature
in suppressing all fluctuations and making phase transitions possible
even for finite lattices.  

As the computed dimensionless surface tension for the square lattice
comes out substantially smaller than 1, we suspect that the critical
end point is not too far off and that one would need to go to very
large $N$ to see the first order transition. A precise estimate would
require a theory of the fluctuations, which we do not have in hand at
this point.  Our computation on the triangular lattice suggests that
one may be better off trying there. Of course in neither case does the
phase transition survive to $N=2$ where one knows from studies of the
Heisenberg model that it isn't there. (We believe that a similar
estimate for the related RP$^{N-1}$ problem will explain the failure
of Sokal and collaborators to observe a first order transition in
simulations for $N$ as large as 8.)

\begin{figure}
\begin{center}
\epsfig{file=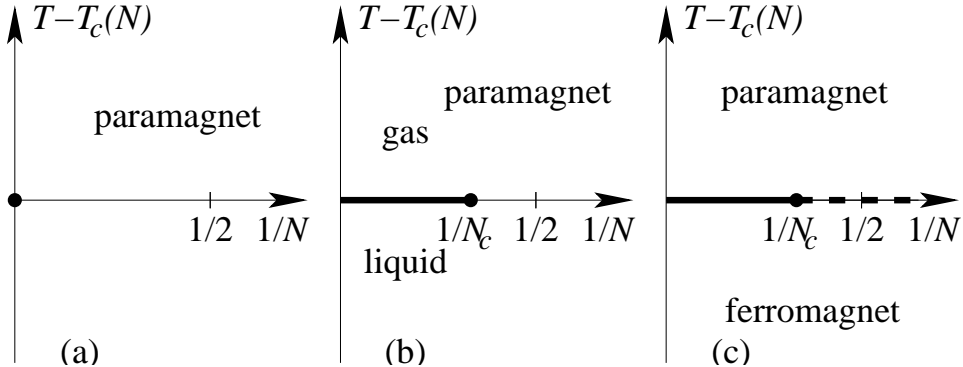,width=5in}
\end{center}
\caption{Suggested phase diagrams of SU($N$) ferromagnets.  First and
second-order transitions are shown as solid and dashed lines,
respectively.}
\label{fig-phases}
\end{figure}

\subsubsection{Second-order cases}
In these cases we conclude that the transition exists {\it only}
at $N=\infty$. Essentially, this situation is the limiting case of the 
one in the previous section where the magnitude of the first order jump 
at $N=\infty$ has gone to zero and hence the critical end point 
has been displaced all the way to $N=\infty$. The important caveat
is that this limit is singular as far as the properties of the
critical point are concerned. The ``critical point'' {\it at}
$N=\infty$ has no fluctuations and has zero correlation length. At
large $N$ there is a ``boundary layer'' where the correlation length
interpolates between a small microscopic value and infinity.

\subsubsection{Transitions in $d \ge 3$}

Starting with $d=3$ it becomes possible to break the $SU(N)$
symmetry. A cursory examination of the $N=\infty$ theory shows
that cases in which the comparison of the paramagnetic solutions
suggests a large jump in $|Q_{ij}|$ end up going directly 
between the seriously disordered paramagnet and the ``ferromagnet''
in which SU$(N)$ is broken. Consequently the transition survives
when $N$ is reduced and one expects that this line of first
order transitions turns into a line of continuous transitions
by the time the known SU(2) cases are reached.

The other new possibility, in $d\ge 4$ is that the inter-paramagnetic 
transition survives for a range of $N$ and then terminates in a line of
confinement/deconfinement transitions. Deciding whether this
is likely is beyond the methods used in this paper and of academic
interest in the study of quantum magnets.


\section{Conclusion}
\label{section-conclusion}

In the foregoing analysis we have established that the infinite $N$
transition between two paramagnetic phases is {\it not} generically,
an artefact of that limit. It {\it is} however a delicate transition,
since it relies on a large entropy from the number of flavors
overpowering the energetics. Consequently, as illustrated by our
surface tension computation, we do not expect it to survive to small
values of $N$. As such, while there is no contradiction between its
existence at large $N$ and the failure to observe it at values of $N$
accesible by other methods, it does mean that the large $N$ finite
temperature phase diagram is not a reliable guide to the SU$(2)$
case---which assumption is the basis of large-$N$ treatments. The same
is true in higher dimensional cases where the infinite $N$ transition
is to the ferromagnetic phase. We have also shown that the violation
of Elitzur's theorem at infinite $N$ is a consequence of a divergence
between the exchange constant and the transition temperature and does
not, in itself, invalidate the infinite $N$ analysis.

\noindent
{\bf Acknowledgements:} We are extremely grateful to Roderich Moessner
for early discussions on this problem and for collaboration on related
work.  It is a pleasure to thank Paul Fendley for collaboration on a
companion paper.  We are also very grateful to David Huse for his
customary enlightening comments on various statistical mechanical
issues discussed in this paper and to Subir Sachdev and Nicholas
Read for sharing their wisdom on large $N$ methods.


\appendix

\end{document}